\documentclass[fleqn,10pt]{wlscirep}

\usepackage{graphicx}
\usepackage{dcolumn}
\usepackage{bm}
\usepackage{booktabs}

\graphicspath{{images/}{plots/}}

\newcommand{\avg}[1]{\ensuremath{\left< #1 \right>}}

\newcommand{\abs}[1]{\ensuremath{\left\vert#1\right\vert}}

\DeclareMathOperator{\Var}{Var}

\DeclareMathOperator{\sgn}{sgn}

\title{Collective effects of the cost of opinion change}
\author[1,*]{Hendrik Schawe}
\author[1,+]{Laura Hern\'{a}ndez}

\affil[1]{Laboratoire de Physique Th\'{e}orique et Mod\'{e}lisation, UMR-8089 CNRS, CY Cergy Paris Universit\'{e}, France}

\affil[*]{hendrik.schawe@cyu.fr}
\affil[+]{laura.hernandez@cyu.fr}

\begin{abstract}
    We study the dynamics of opinion formation in the situation where changing
    opinion involves a cost for the agents. To do so we couple the dynamics of a
    heterogeneous bounded confidence Hegselmann-Krause model with that of the
    resources that the agents invest on each opinion change. The outcomes of the
    dynamics are non-trivial and strongly depend on the different regions of the
    confidence parameter space. In particular, a second order phase transition,
    for which we determine the corresponding critical exponents, is found in the
    region where a re-entrant consensus phase is observed in the heterogeneous
    Hegselmann-Krause model.
    For regions where consensus always exist in the heterogeneous Hegselmann-Krause
    model, the introduction of cost does not lead to a phase transition but just to
    a continuous decrease of the size of the largest opinion cluster. Finally in
    the region where fragmentation is expected in the heterogeneous HK model, the
    introduction of a very small cost paradoxically increases the size of the
    largest opinion cluster.
\end{abstract}

\begin{document}
    % directly from the example
    \flushbottom
    \maketitle
    \thispagestyle{empty}

    \section{Introduction}

    Mathematical models of opinion dynamics aim in general, to understand the
    systemic consequences of a dynamics based on interactions among agents derived
    from Social Influence Theory~\cite{latane1981psychology}. In this framework a
    large variety of models have been studied: discrete---binary or multiple---or
    continuous opinion variables, scalar or vector opinions, mixed or networked
    population~\cite{Castellano2009Statistical, Sirbu2017Opinion}.

    The influence of quenched disorder in the form of idiosyncratic properties of
    the agents has also been studied, like the inclusion of a proportion of stubborn
    or militant agents in the society~\cite{galam2004contrarian,galam2007role,ghaderi2014opinion},
    or the role of the heterogeneity in the confidences that agents hold on each
    other~\cite{Lorenz2010Heterogeneous,Kou2012multi,Pineda2015mass,Han2019Opinion,lorenz2003opinion,Liang2013Opinion,Shang2014agent,Guiyuan2015Opinion}.
    Very recently we have presented a numerically extensive study on the role of
    heterogeneous quenched confidences on the bounded confidence Hegselmann-Krause (HK)
    model~\cite{schawe2020open}, which will be the starting point of the work we present here.

    In most cases, these works include in their models two basic psychological aspects,
    conceptualized by Deutsch and Gerard~\cite{deutsch1955study}, that lead the
    agents to conform to the opinion of others. This property of human interaction
    is often modelled by \emph{homophily}, the tendency of an agent to adapt its opinion to
    conform with its neighbours, and \emph{social influence}, the fact that an agent
    is more influenced by those that are already more alike. One of the great successes
    of the mathematical modelling of opinion formation in terms of a dynamical system was
    precisely to understand how we can observe different opinions coexisting in a society,
    despite the fact that the two considered interactions tend to homogenize
    the opinions of the social actors~\cite{axelrod}.

    This scheme is well adapted to understand the dynamics of opinion formation
    about many subjects of social life. However there are situations where modifying
    its own opinion may involve a particular effort for the agent, to the extent that if
    the agent cannot afford this effort, then it will be impossible for it to evolve.
    Social issues of this kind are those where the change of opinion automatically involves
    a behavioural change. This is not always the case: changing the opinion about the
    literary qualities of an author after a discussion with a well informed friend is one
    thing, but let us consider a subject like climate change. In this case, passing from a
    careless point of view to a more strict position concerning the preservation of the environment,
    does involve an effort to fundamentally change one's behaviour. It can be said that in
    this kind of situation, changing one's opinion bears a \emph{cost}. This cost may be
    strictly economic, like spending more money on locally produced food instead of
    cheap food produced on an area which was, until recently, rain forest.
    % * the car example is probably problematic, since producing a new car will
    % be worse (except maybe for commercial trucks) for the environment than using an old one
    % changing our car to a---more expensive---less polluting one,
    % * the "more time" example seemed a bit abstract, so I merged it with the plane
    % Or it may be just behavioural, like the need of restricting plane travels,
    % or spending more time in dealing with everyday tasks to make them less harmful for
    % the environment.
    Or it may be just behavioural, like using the train instead of travelling by plane,
    which incurs the cost of an increased travel time but is less harmful for
    the environment.

    One of the motivations or studying the role of the cost of opinion change, is to
    try to understand one of the aspects of the theory that Anthony Downs postulated
    to describe the behaviour of public attention face to some issues that can dramatically
    affect social life, known as \emph{Downs' Attention cycle}~\cite{Downs1972}. When a
    society faces a problem that affects at the beginning a small fraction of its members
    but that carries a potential threat to the whole society, according to Downs, the public
    interest towards the problem will undergo five phases:
    \begin{enumerate}
        \item Pre-problem stage. Only a small fraction of the population is aware
            of the problem, either because they are affected or because they are specialists.
        \item Alarmed discovery and euphoric enthusiasm. More and more social actors
            think that the problem should be addressed.
        \item Realizing the cost of significant changes. Social actors realize that a
            big effort (either economic or behavioural) is required to solve the problem.
        \item Gradual decline of intense public interest. Upon realizing of the high
            cost of necessary changes, most actors become discouraged and do not support the
            changes anymore.
        \item The post-problem stage. Although social interest has been re-directed
            to other urgent problems, the global attention devoted to solve the original one
            is still higher than in the pre-problem stage.
    \end{enumerate}

    In this work we concentrate on the notion of cost associated to the opinion change that
    intervenes in phases 3 and 4 of Downs' attention cycle. By introducing a simple model that
    couples the dynamics of a heterogeneous bounded confidence HK model to that of the resources
    of the agents, which decrease as they are invested in sustaining successive opinion changes,
    we want to understand to what extent taking into consideration the cost of opinion change
    modifies the outcomes of this dynamics.

    The article is organized as follows: in Section~\ref{sec:methods} we introduce the model and
    the methods applied, in Section~\ref{sec:results} we describe the results and discuss them in
    corresponding subsections, and finally the conclusions and perspectives are presented in
    Section~\ref{sec:conclusions}.

    \section{Models and Methods}\label{sec:methods}
        In the heterogeneous Hegselmann-Krause model~\cite{krause2000discrete,hegselmann2002opinion}
        each agent $i$ has a continuous dynamical variable, $x_i(t) \in [0, 1]$, representing the
        state of its opinion about a given subject and a quenched variable $\varepsilon_i$ that
        represents its \emph{confidence}. The fundamental assumption of \emph{bounded confidence}
        models is that one agent only interacts with those who already have a similar
        opinion. The confidence $\varepsilon_i$ then determines
        to what extent the opinion of another agent may differ from that of agent $i$ for the
        interaction to take place.
        More formally the set of interaction partners, i.e.\ \emph{neighbours},
        of agent $i$ is
        \begin{align}
            \label{eq:neighbours}
            I(i, \vec{x}) = \left\{ 1 \le j \le n | \abs{x_i - x_j} \le \varepsilon_i \right\}.
        \end{align}
        Note that this definition includes agent $i$ as a neighbour of itself
        and in the case of heterogeneous confidences, i.e.\ $\varepsilon_i \ne \varepsilon_j$,
        the interaction possibility between two agents is not symmetric.

        The dynamics is defined in discrete time. At each time step a
        synchronous update of all agents is performed, in which every agent
        takes the average opinion of all its neighbours
        \begin{align}
            \label{eq:update}
            x_i(t+1) = \frac{1}{N_i(\vec{x}(t))} \sum_{j\in I(i, \vec{x}(t))} x_j(t)
        \end{align}
        where $N_i(\vec{x}(t))$ is the number of neighbours of agent $i$, the cardinality
        of $I(i, \vec{x})$

        This particular model leaves the initial conditions and the confidences
        as free parameters. In this manuscript we will always use random initial
        conditions of opinions, $x_i(0)$, drawn uniformly from $[0, 1]$.
        The most studied case considers that all the agents have the same level of
        confidence, i.e.~$\varepsilon_i = \varepsilon \forall i$. In this case all agents
        converge to a single
        \emph{consensus} opinion for $\varepsilon \gtrsim 0.2$ \cite{hegselmann2002opinion}.
        Below this threshold the agents converge either to a \emph{polarized}
        state of two dominant opinions or \emph{fragment} into many
        opinion clusters.

        As elaborated in the introduction, we extend the heterogeneous HK model by integrating
        the \emph{cost} that the agent has to afford (whatever the origin of such cost is)
        each time to update its opinion. Each agent is endowed with a limited amount of
        resources, $c_i(0)$, assigned to it at the beginning of the simulation.
        The opinion update of Eq.~\eqref{eq:update} takes place as
        as long as the agent has resources to afford the change, otherwise we cap the change to
        the maximum which can be afforded by the agent:
        \begin{align}
            x_i(t+1) = \begin{cases}
                x_i'                                       &\text{\quad if } \eta\abs{x_i(t) - x_i'} \le c_i(t)\\
                x_i(t) + \eta c_i(t) \sgn( x_i' - x_i(t) ) &\text{\quad otherwise }
            \end{cases}
        \end{align}
                      where $x_i'$ is the results of the classical update Eq.~\eqref{eq:update}
        and $\sgn$ is the usual sign function, i.e.\ $1$ for positive
        argument and $-1$ for negative argument. The central parameter is $\eta$,
        which governs how expensive a change of opinion is.
        Additionally, we have to update the available resources of each agent
        in each time step by subtracting an amount proportional to the magnitude of
        the opinion change from its resources
        \begin{align}
            \label{eq:cost_det}
            c_i(t+1) = c_i(t) - \eta\abs{x_i(t) - x_i(t+1)}.
        \end{align}
        Note that using this definition $\eta$ can be compensated with a factor
        in front of the initial resources $c_i(0)$. Thus, we choose here $c_i(0)$
        always such that its mean is $\avg{c_i(0)} = 0.5$ and treat $\eta$ as
        the free parameter without loss of generality. This model
        reduces to the standard heterogeneous HK model for $\eta = 0$.

        Under these rules, an agent who has run out of resources is not able to move in the
        opinion space anymore and it therefore keeps its opinion for the rest of the simulation.
        However, it is still able to influence other agents. We will call it
        \emph{a frozen agent} in the following.

        These dynamical rules lead to a stable \emph{final} state where the agents'
        opinions stop evolving. We require as convergence criterion that the sum of
        all the changes is low enough. Specifically here this condition reads:
        \begin{align}
            \sum_i \abs{x_i(t+1) - x_i(t)} < 10^{-4}.
        \end{align}

        We performed the simulations by the means of the algorithm
        introduced in our previous work~\cite{schawe2020open}, which uses a tree
        data structure to speed up the simulation, while
        preserving correctness within the precision of the data type used to
        represent the opinion of each agent. This allowed us to carry out extensive
        simulations of systems sizes up to $n=32768$ agents and still be able to
        explore the extended phase space given by $(\varepsilon_l, \varepsilon_u)$
        and the cost $\eta$. Our results are issued from a statistical study over
        different realizations of the random initial conditions $x_i(0)$ and
        quenched disorder, $\varepsilon_i$, with $1000$ to $10000$ samples for
        each point of the parameter space $(\varepsilon_l, \varepsilon_u, \eta)$. However,
        the introduction of cost induces much longer convergence times than for the
        $\eta = 0$ case, therefore the very large sizes shown for
        the heterogeneous HK with no cost~\cite{schawe2020open}, are still beyond reach.

        \subsection{Clustering}\label{sec:clustering}
            The main question for the classical HK model is
            whether the society converges to consensus, splits into
            polarization with two opinions or fragments into many more
            opinions. Due to the averaging dynamics and the lack of noise
            in the classical model, the agents will group into very sharp
            clusters of practically the same opinion value, within the
            precision of the used data type (here single precision IEEE 754 floats
            and a tolerance of $10^{-4}$ to generously account for numerical errors),
            which are therefore quite easy to classify. So a good observable of the
            level of consensus is the mean size of the largest cluster.

\begin{figure}[htb]
                \centering
                \includegraphics[scale=1]{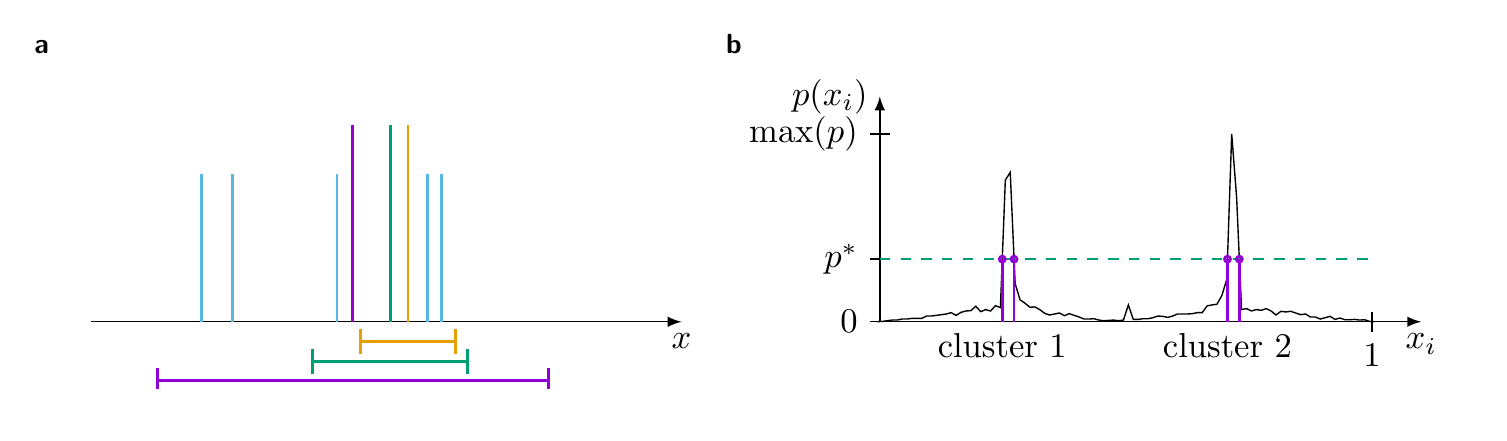}

                \caption{\label{fig:cont_sketch}\label{fig:cluster_sketch}
                    (a) Sketch to visualize the broadening of the clusters. Due to
                    the frozen agents (light blue), agents with different confidences
                    (horizontal bars below the opinion axis) interact with different
                    sets of agents leading
                    to different final positions. Note that each of the non-blue lines
                    represents possibly multiple agents with similar confidence
                    $\varepsilon_i$.
                    (b) Histogram (with 100 bins) of the opinions of agents in
                    the final state for a realization of $n=4096$ agents with a
                    cost of $\eta = 4$. The dashed horizontal line is drawn at
                    $1/3$ of the maximum height. We classify two clusters from
                    the four intersections of the dashed line with the peaks.
                    All agents with a final opinion in the range marked by
                    vertical violet lines are assigned to one cluster.
                }
            \end{figure}

            In contrast to those sharp clusters, the introduction of cost
            will result in broader, quasi-continuous distributions of opinions.
            The fundamental reason for this qualitative change lies in the
            presence of the frozen agents combined with the heterogeneity
            in the confidences. Similar to an effect already observed~\cite{schawe2020open},
            agents of different confidence $\varepsilon_i$ interact with different
            sets of frozen agents, which results in slightly different final
            states. This mechanism is sketched in Fig.~\ref{fig:cont_sketch}(a).
            In the limit of $n\to\infty$, this should actually result
            in a continuous spectrum of final opinions. The higher the fraction
            of frozen agents, the broader the peaks can become. As an example,
            for the broader peaks, a histogram giving the distribution of final opinions
            is shown in Fig.~\ref{fig:cluster_sketch}(b).

            Unfortunately, established methods to classify the clusters of a
            HK model, e.g.\ binning of the opinion space, do not work reliably
            with the relatively broad distributions.
            Therefore, this requires a new robust criterion, to define clusters
            of agents carrying the
            ``same'' opinion. The observable we use is loosely inspired by the
            \emph{full width at half maximum} measurable applied in, e.g.\
            spectroscopy. From each final realization we create a histogram
            to estimate the probability density function $p(x)$ of an agent
            to have opinion $x$ in the final state, as shown in
            Fig.~\ref{fig:cluster_sketch}(b). Then we classify the clusters using
            a threshold of $p^* = \frac{1}{3} \max p(x)$. The clusters are
            defined using the maximal intervals $[x_l, x_u]$, such that
            \begin{align}
                p(x^\prime) \ge p^* \;\forall x^\prime \in [x_l, x_u].
            \end{align}
            For each interval, the set of agents
            $\{ i \, | \, x_l \le x_i \le x_u \}$ constitutes a cluster.
            In other words, each peak larger than the
            threshold is identified as a cluster.
            The choice of the factor $1/3$ to calculate $p^*$ is arbitrary but
            empirically chosen
            to be as small as possible while always being larger than the noise
            floor. Note that this method does not assign each agent to a cluster
            but it is suited as a robust estimate for the size of the largest
            cluster.

        \subsection{Finite-size scaling}\label{sec:fss}

            In the ``Results'' section, we will show that we find evidence of the existence of a
            second order phase transitions when the cost increases. In order to characterize it,
            we use well known finite-size scaling methods~\cite[p.~232]{newman1999monte}.
            For second order phase transitions, one expects the order parameter $S$ to
            obey a \emph{scaling} of the form
            \begin{align}
                \label{eq:scaling}
                S(\eta, n) = n^{-b} f\left[(\eta-\eta_c) n^a\right],
            \end{align}
            where $n$ is the system size, and $\eta$, the cost of each opinion change,
            plays the role of a disordering field, an analogue to the temperature of thermodynamic systems.
            $\eta_c$ is the critical value and $a$ and $b$ are called critical exponents.

            This means that the value of the order
            parameter changes in a region around $\eta_c$ from the typical value
            characterizing one phase to the typical value characterizing the other. This change
            becomes sharper but stays continuous for larger system sizes, i.e.\ the
            region where the order parameter changes becomes smaller.
            The scaling form Eq.~\eqref{eq:scaling} assumes that the region becomes
            smaller as a power law of the system size, such that a value of
            $a < 1$ means that the transition will be sharp in the
            $n \to \infty $ limit.

            $f(\cdot)$ is the \emph{scaling function}, and its most important
            property for our application is that it has no
            explicit dependency on the system size $n$. Thus, if we re-scale
            data measured for different system sizes $n$ with the correct values
            of $a, b$ and $\eta_c$ and plot it as $n^b S(\eta, n)$ over $(\eta-\eta_c) n^a$,
            all data points will collapse on the same curve $f$, in the critical region.

            From a practical point of view, we estimate $a$, $b$ and $\eta_c$, by an
            optimization of the curve collapsing. We use an established
            optimization method~\cite{Houdayer2004low}, which has been successfully applied to
            critical phenomena~\cite{melchert2011dedicated,Norrenbrock2016fragmentation,schawe2016phase,schawe2017ising},
            and also to study scaling functions in other contexts~\cite{mendonca2019asymptotic}.
            The idea is to search in the $(a, b, \eta_c)$ parameter space for
            values such that a quality parameter $Q$ is minimized. This quality
            parameter measures the average distance on the $(n^b S(\eta, n))$-axis
            of each data point from an estimate of the scaling form normalized by
            its standard error. The estimate is obtained
            by a linear fit through the data points of other sizes in that region.
            Conceptually this minimizes therefore the deviation of the data points from each
            other similar to fitting procedures which minimize a $\chi^2$ value.
            Hence, this quality parameter per degree of of freedom has a very
            similar interpretation to a reduced $\chi^2$ value: A quality close
            to $1$ means that all data points lie on average one standard error
            off the estimated curve, which would be the best case one could
            expect. $Q \ll 1$ hints that the standard errors
            of the data are overestimated and $Q \gg 1$ means
            that the data does not fit well to the scaling assumption.
            Since we deal with finite system sizes and and data points are not taken exactly at
            $\eta_c$, values moderately larger than one are expected.
            Additionally this quality parameter $Q$ can be used to estimate an
            error for the values of $a$, $b$ and $\eta_c$, by finding the minimal and
            maximal values for $a$, $b$ and $\eta_c$ at which the quality is $Q+1$,
            i.e.\ where the deviation is one standard error larger.
            We use an implementation of this optimization procedure
            from Ref.~\cite{melchert2009autoscale}.

    \section{Results}\label{sec:results}

\begin{figure*}[htb]
            \centering
            \includegraphics[scale=1]{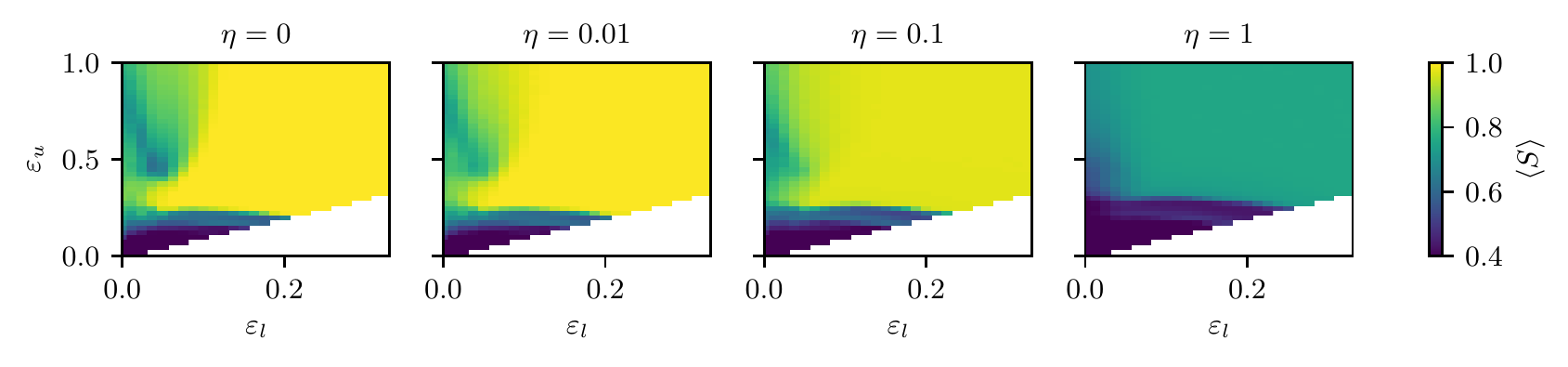}
            \caption{\label{fig:cost_phasediagram}
                Phase diagrams of the consensus for different values of
                the cost $\eta$ in the $(\varepsilon_l, \varepsilon_u)$
                plane determining the heterogeneity. Each image shows $1081$
                points $(\varepsilon_l, \varepsilon_u)$. Each point results from an
                average over $1000$ samples of societies
                of $n=4096$ agents.
                Note the compressed colour range starting at $\avg{S} = 0.4$.
            }
        \end{figure*}

        As an overview of
        the influence of the cost parameter, $\eta$, on societies having agents with confidences
        in different intervals, we show in Fig.~\ref{fig:cost_phasediagram}
        $(\varepsilon_l, \varepsilon_u)$ planes of the phase diagram, for different $\eta$ values.
        The state of the system is measured by the means of the average size of the largest
        cluster $\avg{S}$, according to the clustering criterion introduced before.

        The leftmost panel of Fig.~\ref{fig:cost_phasediagram} shows the previously
        studied \cite{schawe2020open} case of $\eta = 0$ to serve as a baseline to
        observe the influence of cost. The most striking structure, is the
        \emph{re-entrant consensus} region, around $(0.05, 0.3)$.
        Also comparing this case with the results from using a strict
        clustering \cite{schawe2020open} shows that this novel clustering
        is able to detect the same structures, validating the novel method.

        For the very small value of the cost $\eta = 0.01$ the general shape remains,
        but the polarization region (green) at large $\varepsilon_u$ and low $\varepsilon_l$
        becomes less sharp and the structures observed in the $\eta =0$ case, start to smear
        out. Also note that
        at small values of $\varepsilon_l$ but large values of $\varepsilon_u$, consensus actually
        is increased. We will explain this effect later in Sec.~\ref{sec:caseC}.

        For small to intermediate costs $\eta = 0.1$ the re-entrant consensus is destroyed,
        but consensus is still reached when agents with large confidences, $\varepsilon_u$,
        are included.

        Finally, for large cost $\eta = 1$ there is no consensus at all, even above the
        critical value of $\varepsilon_l > 0.2$, where all the agents have a confidence
        larger than the critical value of the homogeneous HK model.

        These observations naturally lead to the following questions:
        Is there a critical value $\eta_c$ above which consensus is suddenly lost?
        Or do we expect to lose consensus smoothly as $\eta_c$ increases?
        And in case where this threshold exists, what is the state for $\eta \ge \eta_c$?
        Is the society polarized, i.e.\ split in
        two clusters, or is there still a majoritarian cluster with a
        homogeneous floor of frozen agents?

\begin{figure}[htb]
            \centering
            \includegraphics[scale=1]{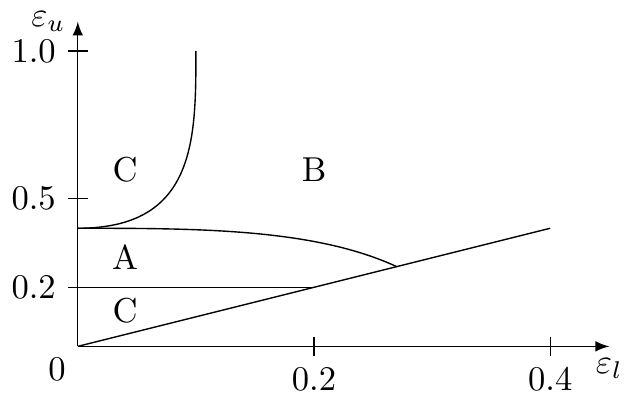}
            \caption{\label{fig:phase_diagram_sketch}
                Regions of the $(\varepsilon_l, \varepsilon_u)$ space showing different behaviour
                with increasing $\eta$, issued from a dozen measurements, i.e.\ the lines
                are only a rough estimate.
            }
        \end{figure}

        Given the rich structure of the phase diagram of the heterogeneous HK model,
        we can split it into three main regions which show a qualitative
        different behaviour. For computational reasons, the precise determination of
        the borders of these regions is out of the scope of this work, we just give here a
        rough schema in Fig.~\ref{fig:phase_diagram_sketch} of the identified regions.
        The following results are issued from simulations done at a dozen points
        $(\varepsilon_l, \varepsilon_u)$ situated in these different regions for
        several values of $\eta$ and multiple system sizes.

        In these 3 regions, we observe three qualitatively
        different classes of behaviour:

        \begin{enumerate}
            \item[A.] The most interesting case, which seems to greatly overlap
                with the region of the re-entrant consensus phase in the case
                of zero cost. There is strong
                consensus at $\eta = 0$ and increasing $\eta$ leads to an
                abrupt loss of consensus towards a polarization. Our analysis in
                Sec.~\ref{sec:phase_transition}, shows that this is a phase
                transition of second order with a critical exponent, which
                seems to be the same over large parts of this region.
                Note that this behaviour is even present for homogeneous cases,
                e.g.\ $\varepsilon = 0.25$.
            \item[B.] Increasing $\eta$ leads to a gradual loss of consensus. The final
                configuration consists typically of one single central
                majoritarian opinion and many agents frozen almost uniformly
                across the opinion space. The form of the curve $\avg{S}(\eta)$,
                i.e.\ the mean size of the largest cluster as a function of the
                cost, shows only very weak size $n$ dependency (corresponding
                to a critical exponent of or close to 0, which means that the
                this is not a sharp phase transition). The formation of
                a second cluster for a polarized configuration like in A, is
                suppressed since agents of one cluster could always see the
                other cluster and would travel towards the central opinion
                or freeze at different points on the way, according to their resources.
                The details of this region are presented in
                Sec.~\ref{sec:caseB}.
            \item[C.] There is no consensus at $\eta = 0$ (see Fig.~\ref{fig:cost_phasediagram}).
                With increasing $\eta$, frozen agents appear, but the opinions stay fragmented
                and no sharp transition can be observed.
                We will show some interesting peculiarities in Sec.~\ref{sec:caseC}.
        \end{enumerate}

        \subsection{Case A: Transition to polarization with increasing cost}\label{sec:phase_transition}
\begin{figure}[hbtp]
                \centering
                \includegraphics[scale=1]{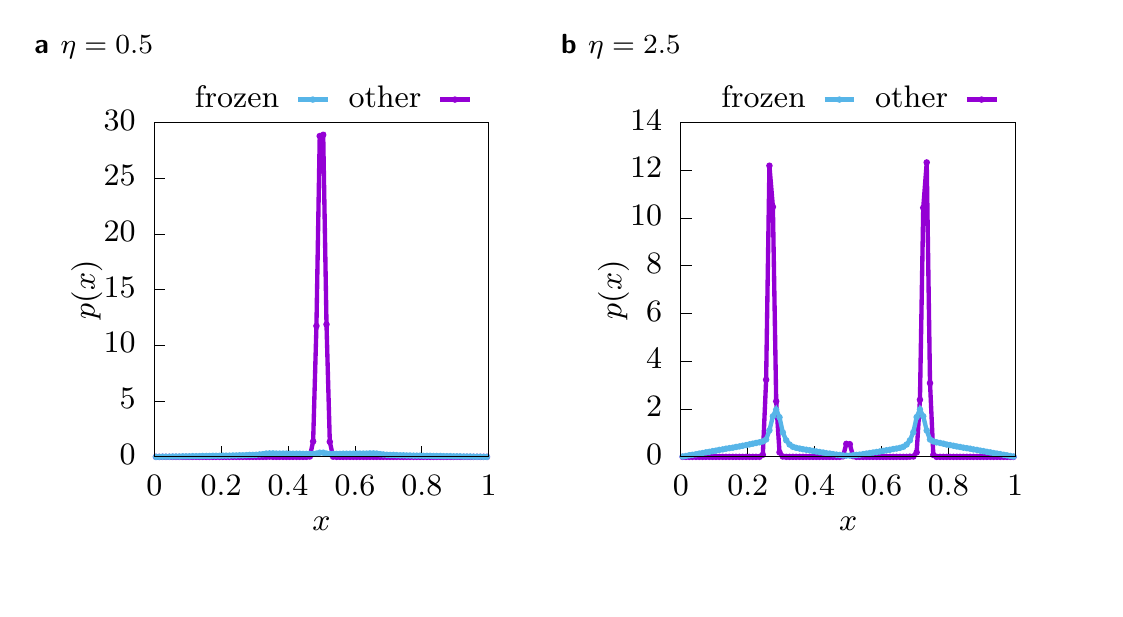}
                \caption{\label{fig:resources}
                    Histogram of the final opinion of agents with resources
                    (violet) and without resources in the final state (light blue).
                    The histogram was collected over $10000$ samples of systems
                    with $n = 16384$ at $(\varepsilon_l, \varepsilon_u) = (0.1, 0.3)$
                    agents with (a) low costs of $\eta = 0.5$ and (b) high
                    costs of $\eta = 2.5$.
                }
            \end{figure}

            First, we will analyse the distribution of final opinions to
            understand the role of the frozen agents in the equilibrium
            configurations of both phases. Figure~\ref{fig:resources} shows
            the final configurations for $(\varepsilon_l, \varepsilon_u) = (0.1, 0.3)$
            and $n = 16384$ averaged over $10000$ samples below and above the
            critical cost $\eta_c = 0.909(6)$ (see below).
            Frozen agents which have spent all their
            resources (light blue) and those which still interact as usual (violet), are shown.
            Figure~\ref{fig:resources}(a) below the critical $\eta_c$ shows a clear
            consensus configuration with a single peak and almost no frozen agents.
            Figure~\ref{fig:resources}(b) far above $\eta_c$, shows a clear
            two-peak structure, indicating polarization. Indeed, even for this
            extremely high cost, most agents
            still have resources in the final state, and even a large fraction of
            the frozen agents, i.e.\ those without resources, are located within the
            bounds of one of the two clusters. The small peak in the centre
            (also visible in the single realization shown in Fig.~\ref{fig:cluster_sketch})
            is caused by agents ending up with a central opinion because, given their
            confidence values, they can interact with both peaks and end up with
            the average opinion of both clusters.

\begin{figure}[hbtp]
                \centering
                \includegraphics[scale=1]{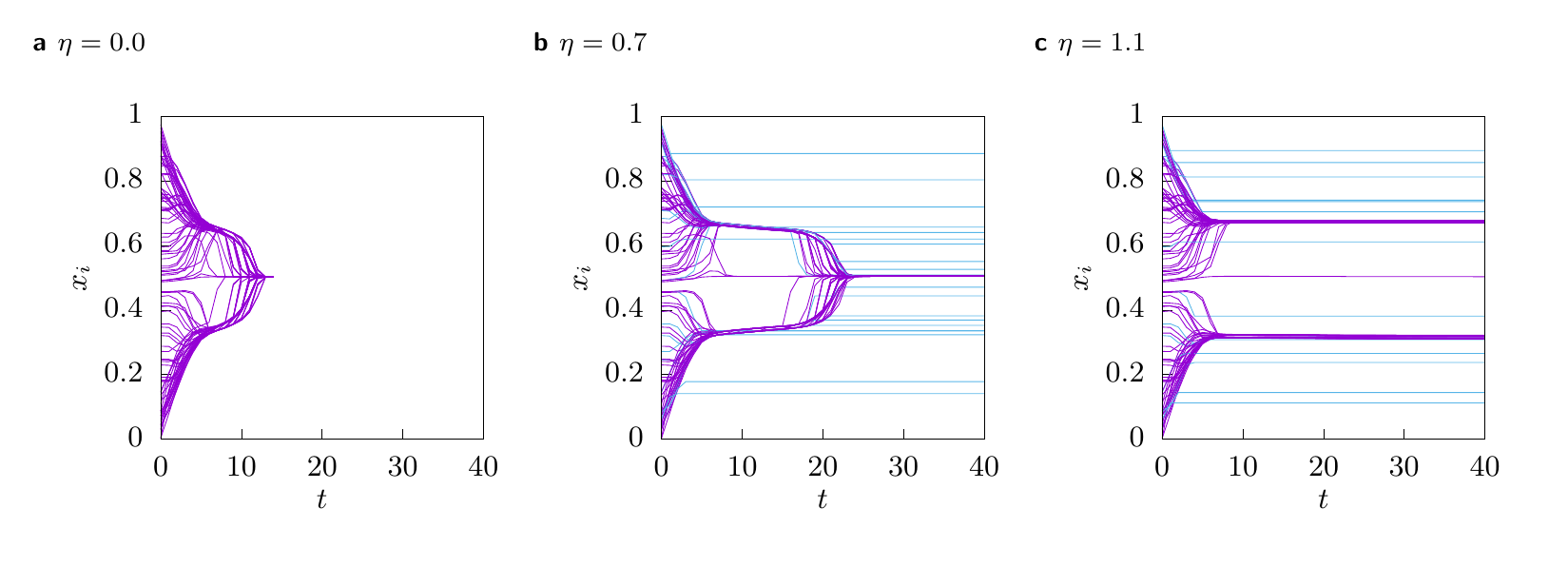}
                \caption{\label{fig:trajectory}
                    Single trajectories of the same system with $n = 16384$
                    agents for different values of the cost: (a) $\eta = 0$,
                    (b) $\eta = 0.7$, (c) $\eta = 1.1$. To improve the visualization
                    only the trajectories of $100$ randomly sampled agents are shown in the figure.
                  }
            \end{figure}

            \paragraph{The dynamics} In Figure~\ref{fig:trajectory} three examples of
            trajectories of the same system at $(\varepsilon_l, \varepsilon_u) = (0.1, 0.3)$
            with the same initial opinions $x_i(0)$, the same confidences
            $\varepsilon_i$ and the same initial resources $c_i(0)$, but with
            different values of the cost $\eta$ are shown. Note that the system
            has $n = 16384$ agents but for clarity of the figures only the
            trajectories of $100$ randomly chosen agents are visualized.
            Figure~\ref{fig:trajectory}(a) shows the dynamics of the system at
            $\eta = 0$. We can observe the `bell' structure~\cite{schawe2020open}
            which facilitates consensus despite the high proportion of very
            closed minded agents. This system converges to total consensus
            after $t=15$ steps. Figure~\ref{fig:trajectory}(b) shows the
            influence of a low cost: Multiple agents freeze outside of the
            `bell', such that the two branches are pulled away from each other
            leading to an increase in the time needed to reach consensus.
            Note that this effect is caused by agents with very low initial
            resources $c_i(0)$ and we will see in Sec.~\ref{sec:c_dist} that
            the exclusion of of agents with initial resources close to zero
            drastically changes the character of the transition.
            With increasing cost more agents freeze outside of the `bell'
            such that the branches are pulled stronger away from the centre,
            preventing consensus to form, as show in Fig.~\ref{fig:trajectory}(c).

            \paragraph{The order parameter} We study the behaviour of the mean relative size of
            the largest cluster $\avg{S}$ as a function of $\eta$. In the
            consensus case, at $\eta = 0$, one expects $\avg{S} = 1$. In the
            polarization phase, one expects $\avg{S} \le 0.5$, since a few
            frozen agents will be scattered outside of both clusters.

\begin{figure}[htb]
                \centering
                \includegraphics[scale=1]{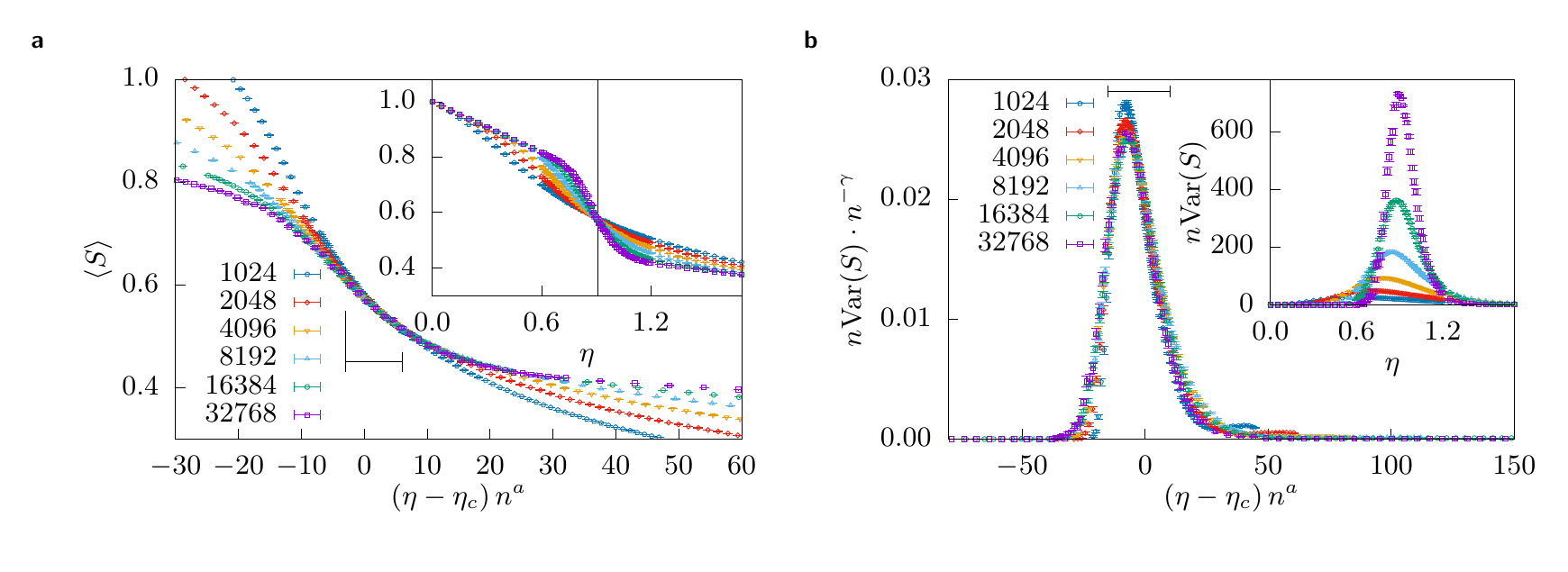}

                \caption{\label{fig:s_collapse}\label{fig:sus_collapse}
                    (a)
                    Inset:
                    The mean size of the largest cluster $\avg{S}$ (cf.~Sec.~\ref{sec:clustering})
                    for multiple system sizes $1024 \le n \le 32768$ as a function
                    of the cost $\eta$ for $(\varepsilon_l, \varepsilon_u) = (0.1, 0.3)$.
                    Each data point is averaged over $10000$ samples.
                    Note that the density of data points is higher close to
                    the transition point and that error bars are smaller
                    than the symbols.
                    Main Plot:
                    The same data but with rescaled axes (cf.~Sec.~\ref{sec:fss}).
                    The black marker shows the range over which the quality of
                    the collapse $Q$ was optimized. The exponent for the vertical
                    axis was fixed at $b=0$, the value for the critical point
                    and the critical exponent $a$ as obtained by the optimization
                    are listed in table~\ref{tab:crit}.
                    (b) The same but for a quantity analogue to the susceptibility
                    $\chi = n \Var(S)$.
                }
            \end{figure}

            The inset of Fig.~\ref{fig:s_collapse}(a) shows $\avg{S}(\eta)$ for a
            selection of system sizes $n$. The curves become steeper for
            larger system sizes.
            This behaviour is typical of second order phase transitions.
            Since our order parameter, $\avg{S}$, is dimensionless, it is expected
            that they cross at the critical value of the control parameter
            $\eta_c$ (cf.~e.g.~percolation or cumulants for magnetic transitions),
            which is indeed the case here.
            Also this means that we do not have to scale the vertical axis, hence $b=0$.

            \paragraph{Scaling behaviour} In order to find critical parameter
            $\eta_c$ and critical exponent $a$, we apply the optimization
            technique described in Sec.~\ref{sec:fss}
            such that curves for all system sizes collapse on the common scaling
            function when the axes are appropriately rescaled. This is shown
            in the main plot of Fig.~\ref{fig:s_collapse}(a).
            Using the values
            $a = 0.45(4)$ and $\eta_c = 0.909(6)$, it is possible to collapse
            all curves on the common scaling function. In the range marked by
            black lines, the curves for $n > 4096$ deviate on average by
            $0.39$ standard errors from each other. Considering the very
            small statistical errors in the data (the error bars are always
            smaller than the symbols), this indicates a very good
            quality of the collapse.

            The concept of universality in the context of phase transitions
            means that the fundamental behaviour of the transition, i.e.\ the
            critical exponents, are robust against details of the model and only
            depend on the symmetries of the interactions or the dimension of the system.
            Here, we see hints of universality in regards to the values of
            $(\varepsilon_l, \varepsilon_u)$, as long as we are still in region A
            of the phase diagram.

            The inset of Fig.~\ref{fig:sus_collapse}(b) shows the behaviour of the fluctuation of
            the order parameter $\chi = n \Var(S)$ (the analogue to the susceptibility
            of magnetic systems),
            which exhibits all characteristics expected at criticality:
            it diverges with system size, its peak moves towards the critical
            point for increasing system sizes and it can be collapsed using
            a second critical exponent $\gamma$ to scale the vertical axis.

            Note that the collapse in the main part of Fig.~\ref{fig:sus_collapse}(b)
            is obtained by an independent optimization over the region $[-15, 10]$ of the rescaled horizontal axis,
            for sizes $n \ge 8192$, yielding $\eta_c = 0.970(6)$, $a = 0.45(1)$ and
            $\gamma = 0.987(5)$ with a quality of $Q = 5.2$.
            % if we use exactly the values of the order parameter: S = 114
            As usual, the collapse of a response function is not as good as for the
            order parameter $\avg{S}$.
            In fact, the quite large quality factor hints that the sizes might not be
            large enough to get a reliable estimate which probably causes the estimated
            critical value to deviate beyond error bars from the one obtained by the
            collapse of the order parameter. Therefore the value obtained from the
            collapse of the order parameter should be considered as an better estimate
            for the actual critical value. Nevertheless, the optimized collapse
            still yields $a$ and $\eta_c$ which deviate less than $10\%$ from the values given in
            table~\ref{tab:crit}. As the deviations are mainly observed for the smallest sizes
            and a convergence to the common curve is visible as size increases, it is
            expected that the results will meet those obtained for $\avg{S}$ in the
            thermodynamic limit.

\begin{figure}[hbtp]
                \centering
                \includegraphics[scale=1]{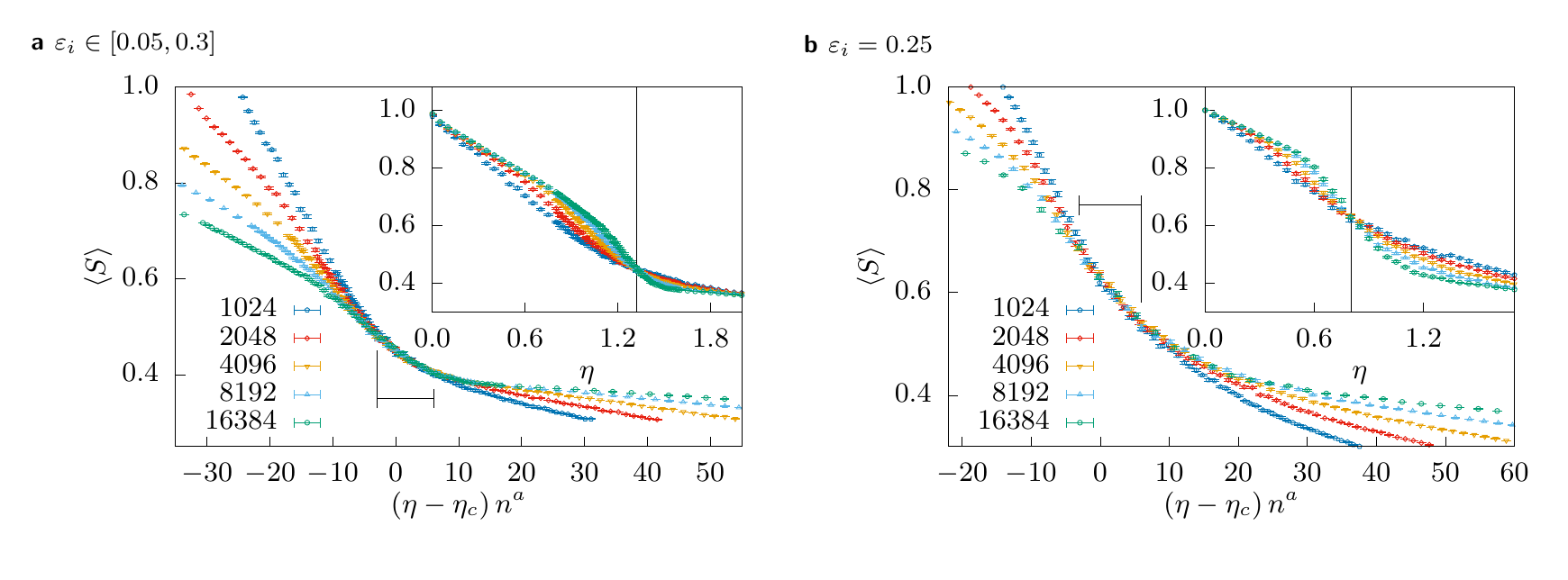}
                \caption{\label{fig:collapse2}
                    Insets:
                    The mean size of the largest cluster $\avg{S}$
                    for multiple system sizes $1024 \le n \le 16384$ as a function
                    of the cost $\eta$ for (a) $(\varepsilon_l, \varepsilon_u) = (0.05, 0.3)$
                    and (b) $\varepsilon_i = 0.25$.
                    Each data point is averaged over $1000$ samples.
                    Main plots:
                    The same data but with rescaled axes.
                    The black marker shows the range over which the quality of
                    the collapse $Q$ was optimized. The exponent for the vertical
                    axis was fixed at $b=0$, the value for the critical point
                    and the critical exponent $a$ as obtained by the optimization
                    are listed in table~\ref{tab:crit}.
                }
            \end{figure}

            Figure~\ref{fig:collapse2} shows the same rescaling for different points
            of region A including the homogeneous case, shown here for $\varepsilon_i = 0.25$.
            The critical point and exponent were determined using the same technique
            as before and all values are compatible within statistical errors.
            Their exact values as well as used range and the quality are shown
            in table~\ref{tab:crit}. In Section~\ref{sec:c_dist} we will explore
            the robustness of this phase transition against changes of the
            distribution of initial resources, $c_i(0)$, and we will show
            a particular case in which this universal behaviour may be destroyed.

            \begin{table}[htb]
                \caption{\label{tab:crit}
                    Critical values $\eta_c$ and exponents $a$ obtained with
                    the optimization approach from Sec.~\ref{sec:fss} (with a
                    quality of $Q$ over the \emph{range} on the rescaled axis)
                    for different value of $(\varepsilon_l, \varepsilon_u)$ and
                    different distributions of the initial resources $c_i(0)$.
                    The critical exponents $a$ are robust
                    to changes in the initial resource distribution as long as
                    there are still agents with very little starting resources.
                    Note that always the mean initial resources $\avg{c_i(0)} = 0.5$.
                }
                \begin{tabular}{rlllll}
                    \toprule
                     & $(\varepsilon_l, \varepsilon_u)$ & \multicolumn{1}{c}{$\eta_c$} & \multicolumn{1}{c}{$a$} & \multicolumn{1}{c}{$Q$} & \multicolumn{1}{c}{range} \\
                    \midrule
                    $c_i(0) \in U[0, 1]$ & $(0.10, 0.30)$ & $0.909(6)$ & $0.45(4)$ & $0.39$ & $[-3, 6]$\\
                    $c_i(0) \in U[0, 1]$ & $(0.05, 0.30)$ & $1.32(1)$  & $0.42(5)$ & $0.89$ & $[-3, 6]$\\
                    $c_i(0) \in U[0, 1]$ & $(0.25, 0.25)$ & $0.80(2)$  & $0.41(6)$ & $1.05$ & $[-3, 6]$\\
                    half-Gaussian        & $(0.10, 0.30)$ & $0.73(1)$  & $0.43(2)$ & $0.76$ & $[-3, 6]$\\
                    $c_i(0) = 1 - (\varepsilon_i - \varepsilon_l) / (\varepsilon_u - \varepsilon_l)$
                                         & $(0.10, 0.30)$ & $0.63(1)$  & $0.43(1)$ & $1.06$ & $[-3, 6]$\\
                    \bottomrule
                \end{tabular}
            \end{table}

            \subsubsection{Influence of the distribution of resources}\label{sec:c_dist}
                The results discussed so far were obtained for a uniform distribution of initial
                resources among the population. This simplifying hypothesis being quite unrealistic,
                we study here whether the obtained results are robust face to different
                distributions of initial resources

\begin{figure}[hbtp]
                    \centering
                    \includegraphics[scale=1]{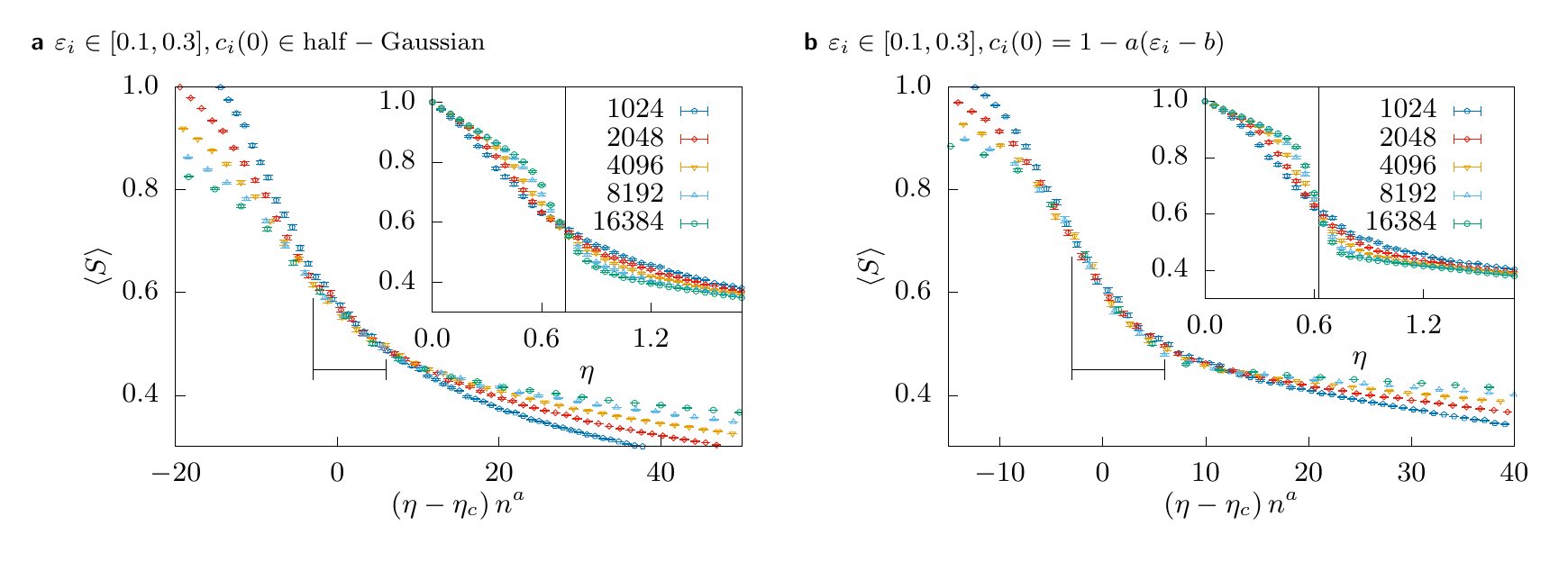}
                    \caption{\label{fig:collapse3}
                        Insets:
                        The mean size of the largest cluster $\avg{S}$
                        for multiple system sizes $1024 \le n \le 16384$ as a function
                        of the cost $\eta$ for $(\varepsilon_l, \varepsilon_u) = (0.1, 0.3)$
                        with initial costs $c_i(0)$ (a) drawn from a half-Gaussian
                        distribution and (b) deterministically chosen as a
                        function of their confidences $c_i(0) = 1 - a (\varepsilon_i - b)$.
                        Each data point is averaged over $1000$ samples.
                        Main plots:
                        The same data but with rescaled axes.
                        The black marker shows the range over which the quality of
                        the collapse $Q$ was optimized. The exponent for the vertical
                        axis was fixed at $b=0$, the value for the critical point
                        and the critical exponent $a$ as obtained by the optimization
                        are listed in table~\ref{tab:crit}.
                    }
                \end{figure}

                Let us consider first, that initial resources $c_i(0)$ are
                distributed according to a half Gaussian
                \begin{align}
                    p(c) = \frac{\sqrt{2}}{\sigma\sqrt{\pi}} \exp\left(- \frac{c^2}{2\sigma^2} \right), \; c>0.
                \end{align}
                In order to keep the system comparable with the previous case, we
                chose $\sigma = \frac{\pi}{2\sqrt{2}}$,
                such that the mean resources per agent $\avg{c_i(0)} = 0.5$, like before.
                So, here agents with very little resources are more probable
                and the richest agents might have $c_i(0) > 1$. Still the
                qualitative behaviour is the same as before: we observe
                in the inset of Fig.~\ref{fig:collapse3}(a) a phase transition
                at a slightly higher critical value. Also the optimization based
                collapse method yields the same universal exponents within error
                bars, which are shown in table~\ref{tab:crit}.

                In the previous cases the resources and the confidences were uncorrelated.
                In order to introduce correlations between them, the simplest hypothesis is:
                \begin{align}
                    c_i(0) = 1 - \alpha (\varepsilon_i - \beta),
                \end{align}
                where again, in order to keep the system comparable to the previous cases,
                $\alpha$ and $\beta$ are chosen such that
                $c_\mathrm{min} = 0$ and $\avg{c_i} = 0.5$. This leads to
                $\beta=\varepsilon_l$, $\alpha = 1 / (\varepsilon_u - \varepsilon_l)$.
                (Note that a similar proportional connection $c_i(0) = \alpha (\varepsilon_i - \beta)$,
                lead to almost identical results as the uniform initial resources.)
                The results show no qualitative deviation from the uncorrelated case, the phase
                transition is shown in the inset of Fig.~\ref{fig:collapse3}(b) and the critical
                exponents are compatible those of table~\ref{tab:crit}. The critical exponents are
                therefore robust face to correlations between confidences and resources.

                Finally, a different qualitative situation occurs for a society without very
                poor agents (those with $c_i(0) \simeq 0 $). Note that this is automatically
                the case for power-law distributions, which generally have a lower cutoff.

\begin{figure}[htb]
                    \centering
                    \includegraphics[scale=1]{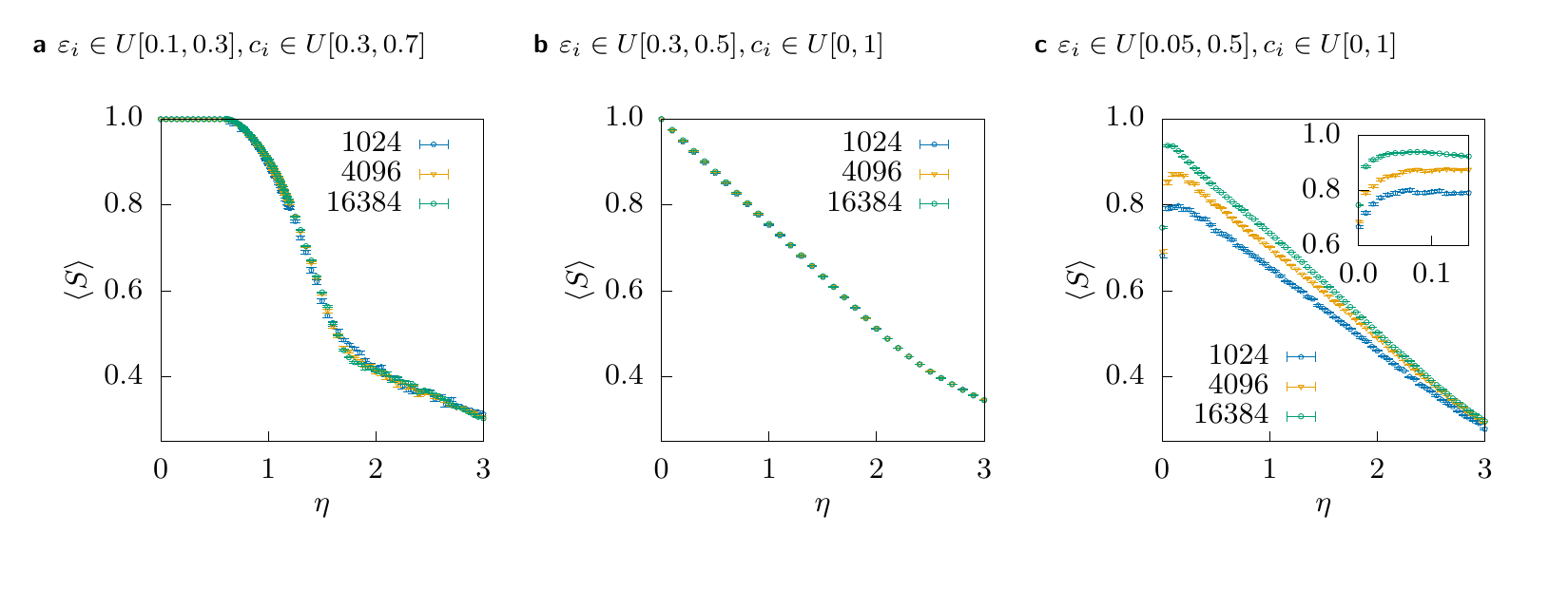}

                    \caption{\label{fig:ex_offset}\label{fig:ex_b}\label{fig:ex_c}
                        Behaviour of the size of the largest cluster with increasing
                        cost for
                        (a) Region A, but with a distribution of the cost,
                        which does not include very poor agents, i.e.\ $c_i$
                        drawn uniformly from $[0.3, 0.7]$.
                        (b) Region B. The decrease of the largest cluster size
                        is independent of the system size, such that we observe
                        even in the infinite system a gradual decline instead of a
                        a sharp transition. Note that for the above cases the data
                        points for different system sizes lie almost on top of each
                        other, such that only the largest size might be visible.
                        (c) Region C. At $\eta = 0$ there is no consensus,
                        i.e.\ $\avg{S}(0) < 1$, the larger the system, the smaller
                        the largest cluster. The inset shows a zoom on small values of $\eta$
                        visualizing that in some fringe cases a small cost can
                        increase the level of consensus in a society
                    }
                \end{figure}

                Fig.~\ref{fig:ex_offset}(a) shows the results for a system where the resources
                are uniformly distributed but within boundaries that prevent the existence of
                very poor agents, here $c_i(0) \in U[0.3, 0.7]$.
                Note that here the upper cutoff is shifted to $0.7$, again in order to keep the
                same mean initial resources $\avg{c_i(0)} = 0.5$ like before.
                The confidences are in the region where a consensus re-entrant phase is observed for
                a system without cost. Interestingly, we observe a decrease of $\avg{S}$ which seems
                independent of the size $n$, suggesting an exponent close to zero, therefore the
                transition never becomes sharp.
                Note that the lack of very poor agents leads to a different dynamics.
                Since all agents have enough resources to travel $\eta c_\mathrm{min}$ in opinion
                space, for moderate values of $\eta$ not a single
                agent will freeze outside of the `bells' that are observed in the trajectories of a
                single realization, in contrast to the case we studied before in Fig.~\ref{fig:trajectory}.
                Agents frozen inside of the bell pull the two branches towards each other, which
                helps consensus. This happens for all the system sizes. When $\eta$ is
                finally large enough to freeze agents outside of the branches,
                a very large fraction of agents is frozen (for $c_i(0) \in U[0.3, 0.7]$ almost $50\%$)
                which leads to a rapid change to polarization, again without relevant size dependence.

        \subsection{Case B: Crossover to fragmentation with increasing cost}\label{sec:caseB}
            Region B of Fig.~\ref{fig:phase_diagram_sketch} corresponds to the region of the
            phase space, where consensus is achieved extremely quickly in the $\eta=0$ case.
            Introducing a cost does not lead to
            polarization as in region A, since the confidences, i.e.\ the
            interaction range of the agents are too large allowing the two polarized
            clusters to interact and converge to consensus. So the final
            configurations in this case consist even for very high cost of a
            single large central cluster and many frozen agents, which are
            spread in the whole opinion space (their density increasing
            linearly towards the central cluster).

            In fact, we can observe in Fig.~\ref{fig:ex_b}(b) that the size of the largest cluster
            decreases linearly in $\eta$ independent of the system size. This
            behaviour is plausible considering that number of agents, with
            $c_i(0) < \eta \abs{0.5 - x_i(0)}$, which do not have enough resources to
            reach the central consensus opinion and freeze outside of the central cluster,
            increases linearly.

            Therefore, in region B the introduction of cost does not lead to a phase
            transition but to a linear decrease of the size of the largest cluster.

        \subsection{Case C: Cost sightly improves consensus in fragmented systems}\label{sec:caseC}
            The last class of different behaviour we observed occurs in region C.
            The main effect we observe in the behaviour of $\avg{S}$ in Fig.~\ref{fig:ex_c}
            is a similar linear decrease with $\eta$ like in region B, but starting
            at lower values, since there is no consensus at $\eta = 0$.

            An interesting effect, we observe is that the introduction of a small cost
            can actually increase the mean size of the largest cluster $\avg{S}$
            with respect to the $\eta=0$ case, paradoxically, increasing consensus.
            Figure~\ref{fig:cost_phasediagram}
            shows this effect, where the top right corner of region C around
            $(0.1, 1.0)$ becomes lighter, i.e.\ more consensus, from $\eta = 0$
            to $\eta = 0.01$. And also the inset of Fig.~\ref{fig:ex_b}(c)
            shows this phenomenon. This contra-intuitive behaviour is caused by frozen
            agents which freeze in between two clusters of the fragmented
            opinion space and act as a bridge: both clusters see the frozen
            agent and are attracted, such that at some point they can interact
            with each other and end up on the same opinion cluster.

    \section{Conclusions}\label{sec:conclusions}
        In this article we investigate the collective effects of the introduction of
        cost in an opinion dynamics model. The stylized model proposed here addresses
        the situations where changing our opinion is not straightforward but requires some
        effort instead. As a consequence the possibility for an agent to change its own
        opinion is conditioned to the resources it has. In particular, in this model, after
        spending all its resources the agent cannot evolve anymore but it is still part of
        the society and its static position may influence the behaviour of others.

        We have coupled the dynamics of the heterogeneous Hegselmann-Krause model
        with that of the resources of the agents, which decrease with time as the agent
        evolves in the opinion space. Intuitively, one expects that the introduction
        of cost would lead, as a trivial consequence, to the reduction of the size of
        the majoritarian opinion group, due to the agents that freeze in different regions
        of the opinion space. Instead, the results of the extensive simulations we have
        performed in the parameter space ($\varepsilon_l, \varepsilon_u, \eta$) show a much
        more complex behaviour depending on the region of the parameter space we are studying.

        The most outstanding result is the way in which the outcomes of the dynamics
        of a society that reaches consensus in the standard HK model is modified. When
        the agents have large confidences, the introduction of cost gradually reduces the
        size of the largest opinion group, as expected. When, on the contrary, the society
        contains agents with intermediate confidences, a second order phase transition is
        observed. By a finite-size study, we have been able to characterize this transitions
        by the critical exponents describing the behaviour of the order parameter and its
        fluctuations and we have observed that they are universal, independent of the details of
        the resource distribution, with one interesting exception, which constitutes the
        second interesting result. If the initial distribution of resources guarantees
        that every agent in the society has a minimum amount of resources, then no phase
        transition is observed and the effect of increasing the cost simply leads again
        to a smooth decrease of the large opinion cluster.

        The interpretation of these results is interesting because it means on one hand,
        that the structure of the population, in terms of the confidences of the agents,
        is relevant in order to evaluate the consequences of the introduction of the cost
        of opinion change: very confident societies will not experience a phase transition
        due to an increase of the cost of opinion change, while societies with intermediate
        confidences will undergo such sharp transition due to an increasing cost.

        On the other hand, in the case of a society with intermediate confidences, an
        initial distribution of resources that does not include very poor agents guarantees
        that there is no sharp transition when the cost increases.

        This study opens different prospective lines, like the role of a networked society,
        the inclusion of the possibility for the agent to choose whether to change its
        opinion according to the amount of its remaining resources or the coupling of
        the cost and the opinions, breaking the equivalence of the opinions of this model
        by making some of them more costly than others.

    \bibliography{lit}

    \section*{Acknowledgements}
        This work was supported by the OpLaDyn grant obtained in the 4th round
        of the Trans-Atlantic platform Digging into Data Challenge (2016-147 ANR OPLADYN TAP-DD2016)
        and Labex MME-DII (Grant No. ANR reference 11-LABEX-0023).

\end{document}